# Universal Spike Classifier

Muhammad Saif-ur-Rehman[1], Robin Lienkämper[1], Yaroslav Parpaley[1], Jörg Wellmer[1], Charles Liu[2], Brian Lee[2], Spencer Kellis[3], Richard Andersen[3], Ioannis Iossifidis[4], Tobias Glasmachers[5], Christian Klaes[1]

[1]Department of Neurosurgery, University Hospital, Ruhr-University Bochum, Germany; [2]Neuroresotoration Center and Department of Neurosurgery and Neurology, University of Southern California, U.S.A; [3]Division of Biology and Biomedical Engineering, CALTECH, U.S.A; [4]Institute of Informatics, University of Applied Sciences, Bottrop, Germany; [5]Institute of Neuroinformatic, Ruhr-University, Bochum, Germany

## Abstract

In electrophysiology, microelectrodes are the primary source for recording neural data of single neurons (single unit activity). These Microelectrodes can be implanted individually, or in the form of microelectrode arrays, consisting of hundreds of electrodes. During recording, some channels capture the activity of neurons, which is usually contaminated with noise and/or artifacts. Another considerable fraction of channels does not record any neural data, but only external artifacts and/or noise. Furthermore, some units get lost from a channel over time, or new units appear on a channel during the recording. Therefore, an automatic identification and continuous monitoring of channels containing neural data is of great significance and can accelerate the process of analysis, e.g. automatic selection of meaningful channels during offline and online spike sorting. Another important aspect is the selection of meaningful channels during online decoding in brain-computer interface applications, where threshold crossing events are usually considered for feature extraction, even though they don't necessarily correspond to neural events. Here, we propose a novel algorithm based on newly proposed way of feature vector extraction and supervised deep learning algorithm: a universal spike classifier (USC). The USC enables us to address both above-raised issues. The USC uses a state-of-the-art deep convolutional neural networks (CNNs) architecture. It takes a batch of raw waveforms as input, propagates it through multiple layered structures, and finally classifies it as a channel containing neural spike data or only artifacts/noise. We have trained the model of USC on data recorded from a single tetraplegic patient with two Utah arrays implanted in different areas of the brain. The trained model was then evaluated on data collected from six epileptic patients implanted with depth electrodes and two tetraplegic patients implanted with two Utah arrays, separately. The implanted electrodes targeted different areas of the brain. The test accuracy was 97.20% on 1.56 million hand labeled test inputs, collected from all eight patients. The results demonstrate that the USC generalizes not only to the new data, but also to brain areas, subjects, and electrodes not used for training.

# 1. Introduction

The human brain is a complex system containing approximately 100 billion neurons (Herculano-Houzel). Neurons communicate by propagating action potentials (Hodgkin and Huxley), also referred as spikes. The spikes generated by individual neurons (sometimes called 'units') can be recorded with the help of microelectrodes (Kita and Wightman). State of the art technological development in microelectronics has allowed the fabrication of tiny but dense microelectrode arrays, containing hundreds of channels (Berényi, Somogyvári and Nagy; Spira and Hai; Lambacher, Vitzthum and Zeitler; Frey, Egert and Heer). As a result, the activities of several thousand neurons can be recorded simultaneously. Generally, each implanted microelectrode records the action potential generated by more than one neuron as well as artifacts in its vicinity, and thus this phenomenon is referred multi-unit activity (MUA). Sometimes, to do further analysis, it is necessary to isolate the activity of each recorded neuron of the electrode, which is also referred as single-unit activity (SUA).The first step to separate the activity of each unit is manual or semi-automatic spike sorting  (Gibson, Judy and Marković; Pachitariu, Steinmetz and Kadir; Matthews BA; Abeles and Goldstein), which includes detecting spike events and then assigning those detected events to specific neurons (SUA). However, because of human curation in this process of spike sorting, the computational methods to interpret the recorded data by new generation probes do not match technological evolution in recording devices. Therefore, in this study, we tried to minimize the role of human intervention during the process of interpretation of neural data.

Most existing spike sorting algorithms use band-pass filtering, spatial whitening and threshold crossing before qualifying an incoming waveform as an event. Finally, they apply clustering on qualified events. This generally involves at least one or more manual processing step (Einevoll GT; Lewicki; Olivier Marre). Moreover, It was shown in previous studies that a considerable fraction of dense implanted microelectrode arrays does not records any neural data, but only external artifacts with high amplitudes and/or noise (Klaes, Kellis and Afalo; Rey, Pedreira and Quiroga; N.Hill, B.Mehta and Kleinfeld; Lewicki). Human involvement in spike sorting can be reduced by automatically identifying and discarding meaningless channels at the first stage before any further analysis. However, the background noise is composed of several complex signals, including external artifacts and noise generated by surrounding electrical components (Einevoll GT; S Lewicki). Hence, the absence of a simple noise model presents a big hurdle to automatically and universally discarding meaningless channels  (Chung JE).

In the last years it has been demonstrated that powerful models can be learned with the help of huge amounts of labelled data and deep artificial neural network architectures (Krizhevsky, Sutskever and Hinton). A specific architecture, convolutional neural networks (CNNs) (LeCun, Bottou and Bengio) with the combination of large labelled datasets (Deng, Dong and Socher; Stallkamp, Schlipsing and Salmen) has transformed the field of computer vision and provided many state-of-the-art results for image classification, object detection and tracking (Guo, Dong and Li). In this study, we use the same approach to solve the problem of discarding channels that do not contain spikes. After collecting and labeling a large amount of spike data, we have successfully trained a deep neural network that enables us to not only to detect and but also to track the channels containing neural data. As a result, noise channels can be discarded before doing further analysis. Our system acts as a universal spike classifier, which can also be used in online analysis. By universal we mean that the trained model can be employed to detect and track the channels containing neural data recorded from different subjects, across different brain areas and even with different types of recording hardware, without any additional preparation. To support this claim, we have evaluated the universal spike classifier on the data recorded from eight different subjects of different age and genders, recorded using different types of microelectrodes across different areas of the brain.

The presented method is based on supervised machine learning, which means ground-truth labels are required to define the cost function (Kotsiantis). It is difficult even for an expert neuroscientist to judge a single event without the context of other events. Here, we introduce a new way of labelling that considers a batch of waveforms from the same channel, instead of a single waveform, to construct a feature vector. This provides the classifier with context for learning and decision making. Our data set contains 1.56 million labelled feature vectors. By mimicking the way humans sort spikes, the USC adapt the quality of generalization. Thus, provides the overall classification accuracy of 97.20%.

The USC can be employed to select meaningful channels during online decoding for brain-computer interfaces, where previously unsorted action potentials were used to extract feature vectors (Klaes, Kellis and Afalo; Sonia Todorova; W Fraser, M Chase and Whitford; Koyama, Chase and Whitford; B Schwartz). Such feature vectors also include the threshold crossing events of the channels which do not contain neural data. In contrast, USC discards all the channels that do not contain neural data. Thus, it allows to consider only those channels where at least minimum number of neural spikes are present. Furthermore, the USC can be applied to the last step of spike sorting, where the user must curate to accept or reject the clusters (Hill, Mehta and Kleinfeld; Kadir SN; Rossant C). With the help of USC, now it is also possible to automatically detect noise and unit clusters. Thus, the human curation is not required anymore at this step.

## 2. Materials and Methods

### 2.1. Approvals

We have considered two kind of patients for this study, tetraplegics patients and epilepsy patients. Tetraplegics subjects were recruited for two different brain-computer interface studies (Klaes, Kellis and Afalo) and were implanted Utah array. These studies were by the approved by the institutional review boards at California Institute of Technology (CALTECH), Rancho Los Amigo, and the University of Southern California (USC), Los Angeles. Further approval details are available in (Klaes, Kellis and Afalo; Aflalo, S and Klaes). Epilepsy patients were regular patients at Knapschafts-krankenhaus, Bochum and they have voluntarily participated int this study.

### 2.2. Implantation information

We have collected data from a total of eight patients (seven males, one female), aged 20 to 63 years. Six of the patients were implanted with microelectrodes in preparation for epilepsy surgery using a Behnke-Fried configuration (Fried, CL and NT). The microelectrodes were coupled in a group of 8 or 16 individual microwires with platinum coated tips. The other two patients were tetraplegics recruited for a brain-computer interface study and were implanted with microelectrode arrays (Utah-Array) (Klaes, Kellis and Afalo; Aflalo, S and Klaes). A single Utah array consists of 100 micro-electrodes arranged in a 10 x 10 grid. The placement of the Utah array was based on a functional magnetic resonance imaging (fMRI) task, the patients performed before implantation, details of the array placement and surgery are described in (Aflalo, S and Klaes; Klaes, Kellis and Afalo). The Utah array electrodes were 1-1.5 mm long and presumably recorded signals from cortical layer 5 (Aflalo, S and Klaes; Klaes, Kellis and Afalo). The array electrode had platinum-coated tips and were spaced 400 μm apart. Further, implantation details are shown in Table 1.

| Subject ID | Sex | Age(Years) | Number of Recordings | Number of Implanted Electrode |
|---|---|---|---|---|
| U1 | Male | 32 | 90 | 192 (2-Utah Array) |
| U2 | Male | 55 | 40 | 192 (2-Utah Array) |
| M1 | Male | 23 | 1 | 16 (Micro-wires) |
| M2 | Male | 63 | 1 | 16 (Micro-wires) |
| M3 | Male | 20 | 1 | 16 (Micro-wires) |
| M4 | Male | 57 | 1 | 08 (Micro-wires) |
| M5 | Female | 52 | 1 | 08 (Micro-wires) |
| M6 | Male | 55 | 1 | 16 (Micro-wires) |

*Table 1: Personal information with implantation details*

## 2.3. Behavioral setup

Recording of meaningful data for spike classification does not require the patients to perform a specific task. However, to collect a versatile dataset, the subjects were engaged in different behavioral tasks. More information about the behavioral task of subject U1 and U2 can be found in previous studies in which the same subjects have participated (Aflalo, S and Klaes; Klaes, Kellis and Afalo). The epilepsy patients performed a reaching task in a virtual reality environment, programmed in Unity 3D (Unity Technologies, San Francisco, CA, USA) and using the HTC Vive virtual reality system, or remained idle during recording.

## 2.4. Data Collection

In the group of tetraplegic subjects, data was collected over a period of 2 years in 2-4 study sessions per week. From the group of epilepsy patients, the data was gathered in 12 months with one study session per subject. The data was recorded using a neural signal processor (Blackrock Microsystems). The raw data was first amplified and then digitized at a sampling rate of 30 kHz. Spike candidates (events) were extracted with a thresholding procedure (Lewicki). The threshold for waveform detection was set to -4.5 times the root-mean-square of the high-pass filtered, full-bandwidth signal with cutoff frequency 250 Hz. Similar settings were used in (Klaes, Kellis and Afalo) to select the meaningful channel and extract feature vectors (unsorted threshold crossing) for online decoding. Each detected waveform consists of 48 samples and represents a time duration of 1.6 ms, containing the 15 samples before and 32 samples after the threshold crossing event. The value at each sampling interval is the corresponding amplitude represented in micro-volts.

## 2.5. Data labeling

We cast the problem of spike classification as a supervised learning task. This means that ground-truth labels were required to train a machine learning model. In a single recording session, a single channel records hundred and sometimes thousands of unlabeled waveforms. Moreover, we have considered eight subjects and 136 recording sessions, this resulted 31.21 million unlabeled waveforms, which may correspond to neural events (action potentials) or any other external events (e.g. muscle activity, noise). We labeled the data in a semi-automatic method consisting of the following steps: Firstly, we applied principal component analysis (PCA) on all detected waveforms of a channel. Secondly, we visualized the first two principal components of a subset of waveforms (see Figure 1 (a2, b2)), which capture the direction of highest variability in the data. After visual inspection, we have employed a Gaussian mixture model (GMM) on the datapoints in PCA space to assign them to clusters, as shown in Figure 1 (a2, b2). The number of clusters with their corresponding centroids (initial points) were defined manually. However, it has also been witnessed in some cases, semi-automatic labeling provide unsatisfactory results. So, in such cases after visual inspection of the clustered points, the waveforms were entirely manually labeled.

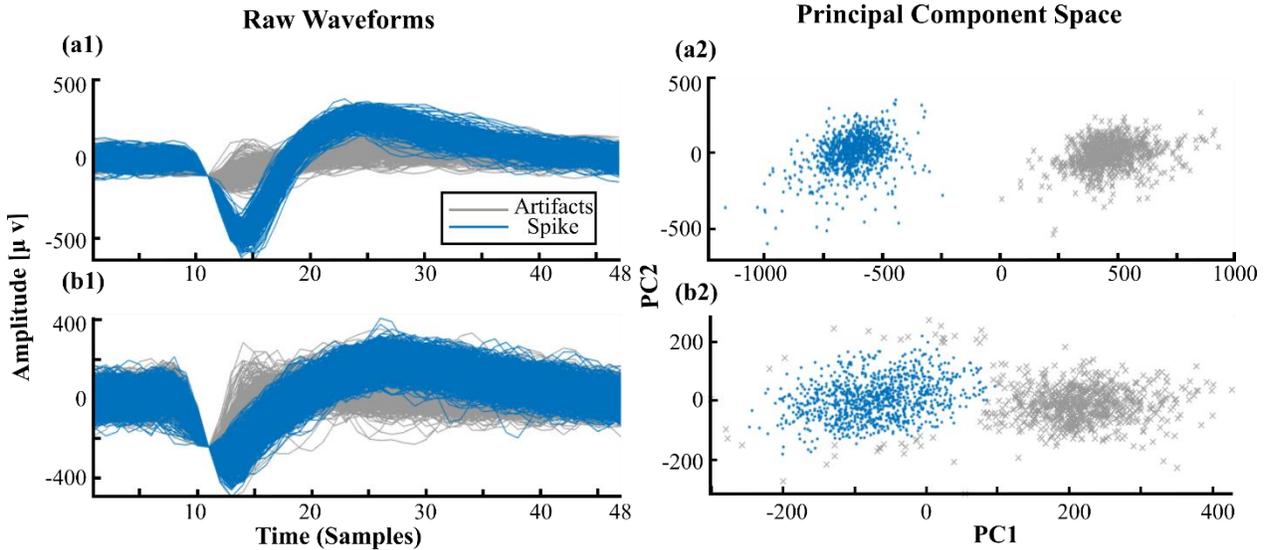

*Figure 1: **The process of labeling of data for two different kinds of channels.** (a1, a2) a1: shows the labeled waveforms of a channel recorded during one recording session, a2: shows the PCA space of the same channel and the result of GMM, used for clustering, the two clusters were easily discriminable. (b1, b2) b1: shows the labeled waveform of another channel recorded during similar recording session as a1, b2: shows the PCA space of the selected channel along with the result of GMM, whereas hand labeling performed along the boundaries of spikes cluster and artifact cluster.*

## 2.6. Batch size

It is hard even for expert neuroscientists to classify a single event as a spike (neural data) or as an artifact in the absence multiple events, e.g., classification of the events shown in Figure 2 (a1, b1). However, when the batch of waveforms of the same channels are considered, simultaneously (see Figure 2(a2, b2)). This process of classification of events becomes relatively trivial because of possibility of contextual learning. Here, we tried to replicate the way humans sort spikes by including context into our feature vectors. Therefore, a single feature vector is constructed by concatenating a batch of waveforms representing multiple events (see Figure 2(a2, b2)), regardless of their category, thus, enabling the universal spike classifier to aggregate the statistics of the inputs in a better way.

Concretely, a single feature vector $x$ is constructed by concatenating a batch of $b$ waveforms of length $w$ (in samples) together, resulting in a vector $x \epsilon R^{b \times w}$. The batch size $b$ is always a positive integer ($b \epsilon N^+$) and is considered a hyperparameter

We have labeled a feature vector $x$ as a spike if at least one of the concatenated events was labeled as a spike during the labeling. Alternatively, if all the concatenated events represent artifacts, then $x$ was labeled as an artifact. For example, if the $b$ equals 20, then to construct a single $x$, twenty successive events of a channel were concatenated together( see Figure 2(a2, b2)), and if any of the twenty events labeled as a spike then, the $x$ was labeled as a spike. Alternatively, if all of twenty individual waveforms were labeled as artifacts, then the $x$ was labeled as an artifact.

Here, we only try to classify which batch of the waveforms contain spikes, and not how many and which of the waveforms in batch represents spikes. This would be the next step and will be referred as universal spike detector (USD).

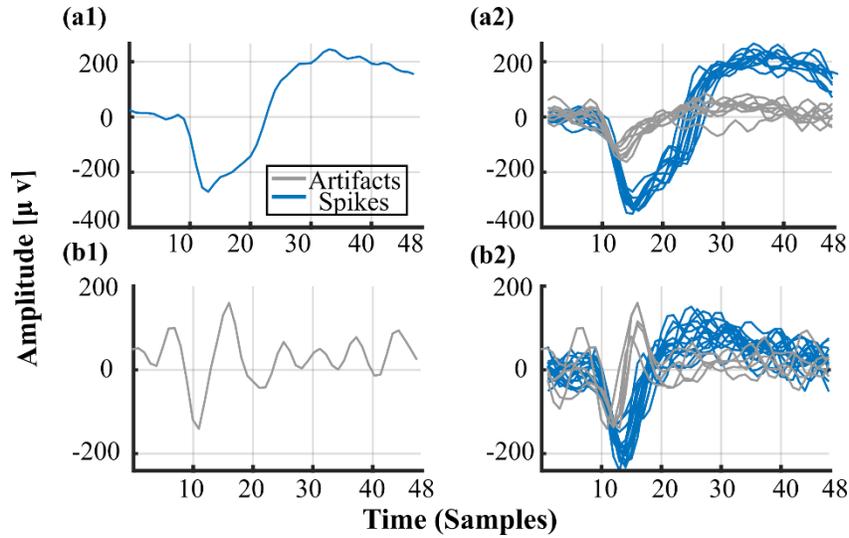

*Figure 2: **Process of construction of feature vector by concatenating the batch of waveforms (Contextual learning).** (a1, a2) a1: shows the individual waveform of the channel representing neural data, a2: shows the 20 concatenated waveforms the same channel representing neural data (spikes) and external artifacts, simultaneously. (b1, b2) b1: shows the individual waveform of the channel resenting external artifact, b2: shows the 20 concatenated waveforms the same channel representing neural data (spikes) and external artifacts, simultaneously.*

### 2.7. Data distribution for training algorithm and evaluation of generalization

The complete dataset contains 31.21 million labeled waveforms. Using a batch size of $b = 20$, that yields 1.56 million labeled feature vectors. The total number of feature vectors resulting from every single individual during recording sessions are shown in *Table 2*.

| Patient Id | No. of Feature Vectors |
|---|---|
| U1 | 756802 |
| U2 | 436898 |
| M1 | 3980 |
| M2 | 5251 |
| M3 | 874 |
| M4 | 13589 |
| M5 | 342951 |
| M6 | 591 |

*Table 2: Number of feature vectors per patient*

In machine learning, generalization is the most significant quality of the algorithm. That means the evaluation performance of the algorithm on unseen data plays a pivotal role. Therefore, for training the algorithm, we have used a small subset of data, compiled from the first six recording sessions of a subject

"U1". Afterwards, we have evaluated the generalization quality of the trained model on the data of all eight subject and all 136 recording sessions.

Figure 3 illustrates the distribution of the labeled feature vectors of both classes within the training set. Class spikes contains feature vectors with a varying number of events representing artifacts, starting from no-contamination (0 artifacts) to maximally-contaminated (19 artifacts), whereas the feature vectors representing class artifact contain events that exclusively represent artifact/ noise. In the available dataset, the spike class provided only 37% of feature vectors (15024), while the artifacts class holds the remaining 63% of the feature vectors (25633). To avoid biases while training the algorithm, we have prepared a more balanced dataset by performing subsampling on the data of both classes. We have selected a dataset $D = \{(x^1, y^1), ... ... ..., (x^N, y^N)\}$ with $N = 30000$ labeled examples, where $x^i$ refers to $i_{th}$ feature vector and $y^i$ to the corresponding class label. From each class, 15000 feature vectors were selected. The feature vector from each class were selected randomly.

We have then sliced the dataset $D$ into a training set $D_{tr}$ containing 70% percent of the data and a validation set $D_{va}$ consisting of the remaining 30%. $D_{tr}$ was used to optimize the parameters of the machine learning model and the hyperparameters of the employed optimization algorithm during training. $D_{va}$ was used to evaluate how well the machine learning model performs on unseen data during training. This validation error was used as a stopping criterion for the training process. The process of training terminated, if the validation error stop decreasing or remain same for six consecutive epochs.

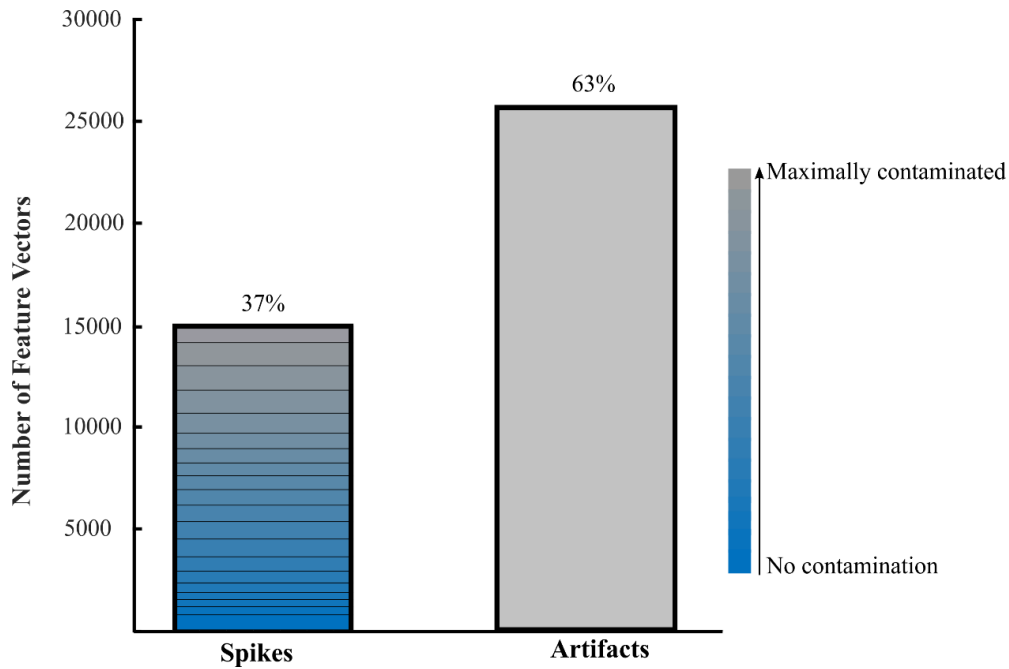

*Figure 3: **Distribution of training data and construction of the feature vector from the batch of waveforms, Subject id: U1, Number of sessions: 6**. In this example, with batch size=20, 20 waveforms were concatenated to get a single feature vector. The range of class spikes feature vectors starts from no contamination, where every single concatenated waveform represents spike event, and ends at maximally contaminated, where 19 of the concatenated waveforms represent artifact events and only one waveform represents spike event. The feature vectors representing class artifacts were harnessed by concatenating the waveforms that exclusively represent artifacts events. 37% of data represents feature vector of class spike and remaining 63% of data represents feature vectors of class artifacts.*

## 2.8. The universal spike classification algorithm

The machine learning model was trained on $D_{tr}$, with the goal to predict the correct label $y_i$ for each feature vector $x_i$ using the output of a learnable parametric decoder $g(x_i; \Theta): x_i \in R^{b \times w} \to y^i$ by learning the parameters Θ, iteratively from $D_{tr}$.

The universal spikes classification algorithm is based on the standard deep architecture of convolutional neural networks (CNNs) (LeCun, Bottou and Bengio). The algorithm followed the standard end-to-end machine learning pipeline. End-to-end learning is decomposed in two parts: The first part maps the raw feature space $x_i$ into the more meaningful feature space $\Phi(x_i; \Theta_\Phi)$ with the learnable parameter matrices $\Theta_\Phi$, resulting features space is more distinguishable. The second part consist of a classifier $f$ with the parameter matrix $\Theta_f$, which maps the feature space $\Phi$ into decision space $g$. The parameters $\Theta_\Phi$ and $\Theta_f$ were learned simultaneously using $D_{tr}$ by iteratively minimizing a single cost function.

### 2.8.1. Input Representation and the Architecture of USC

We have represented the input as a 2-D array with the number of time steps as width and the batch size as the height (shown in Figure 4). To classify the given raw input, we have employed the standard architecture of CNNs used in computer vision task, as explained in (Krizhevsky, Sutskever and Hinton, ImageNet Classification with Deep Convolutional Neural Networks; Guo, Dong and Li, Simple convolutional neural network on image classification). This generic architecture of CNNs is used to extract a wide range of features, which are not restricted to some specific feature type.

The USC contains four convolutional layer and three pooling layers, followed by a fully connected neural network with one hidden layer and a Softmax classifier (Kaibo Duan) as an output layer (see Figure 4 ).

During forward propagation, each filter at each layer is convolved across width and height of the input volume, and then slide with stride=1 over the width and height of the input volume. The resultant will be 2-dimesional convolved feature maps. These feature maps are then further processed through non-linear activation maps. We used Rectified Linear Units (ReLUs) $f(x) = max(x, 0)$ as an activation function (Nair and Hinton).

The first convolutional layer performed convolution across time and the second layer across space (batch size). The preceding convolutional layers perform convolution across time as shown in Figure 4. The size and number of filters at each convolutional layer is mentioned in Figure 4.

Except 1st convolutional layer, each convolutional layer is followed by pooling layer. We used max pooling to down-sample the convolved feature map and to extract more abstract features. The size of each pooling window has height 1 and width 2 in architecture of USC defined in Figure 4.

We padded zeros across the width of the input volume. The zero padding was also added across the width before performing downsampling at conv3 and conv4 in Figure 4.

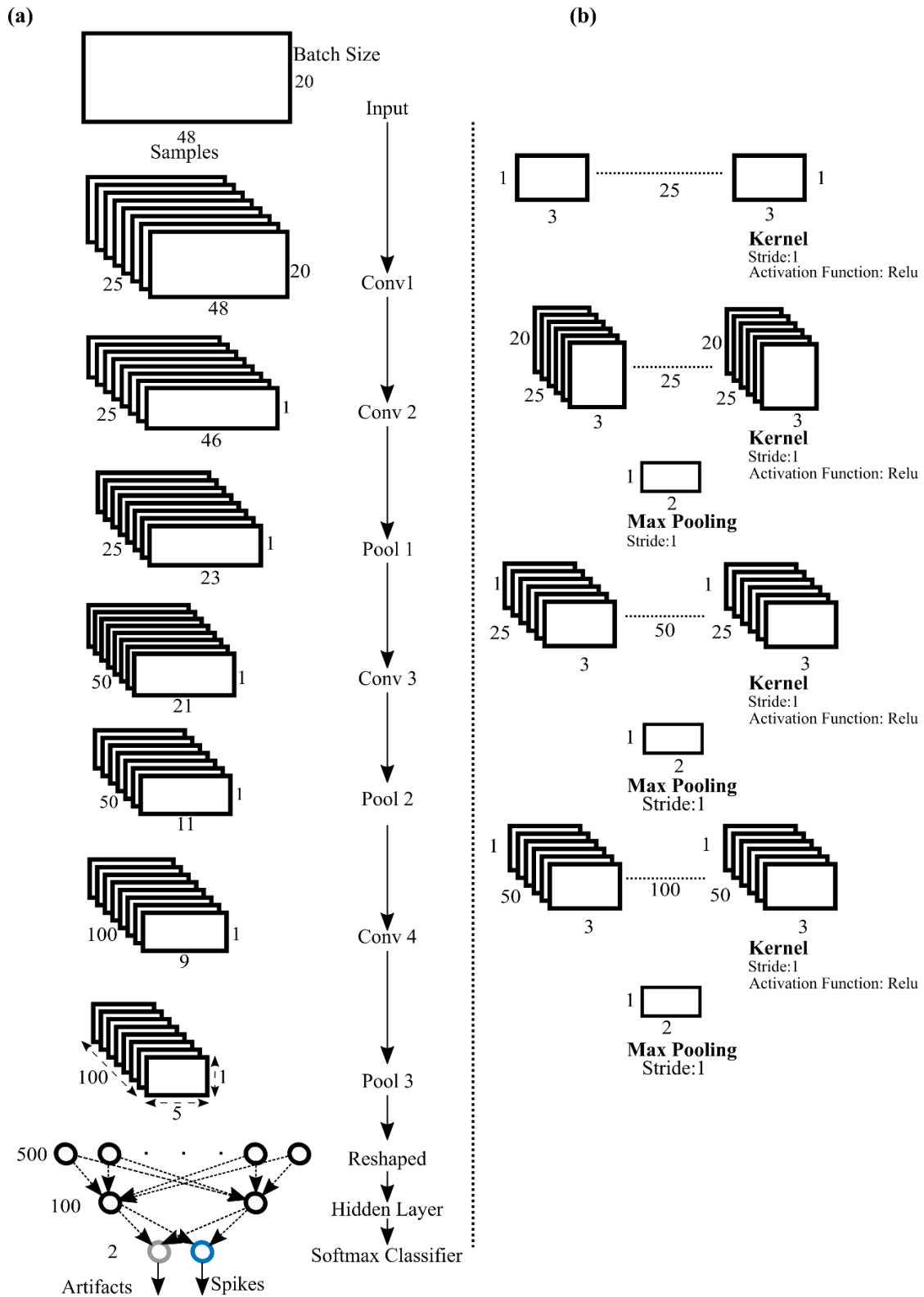

*Figure 4: Architecture of USC. (a) shows the process of mapping input space into decision space. The input is convolved with layer of kernels to get convolved features map. pooling layer down-samples the convolved feature maps. Output of the proceeding layer becomes the input of the following layer. Finally, the Softmax classifier is used to classify the given input. (b) shows the size, stride, and number of kernels along with employed activation function to get the convolved feature map. Max pooling is used to down-sample the convolved feature map, size and stride of the pooling is also given in the figure.*

### 2.8.2. Regularization techniques and optimization algorithm

We have used batch normalization as a regularization technique, which standardizes intermediate outputs of the USC to zero mean and unit variance, for the training example equals to mini-batch size (Ioffe and Szegedy). This helps the employed optimization algorithm during training by keeping inputs closer to normal distribution. The batch normalization is applied to the output of the convolutional layer before nonlinearity (ReLUs), as suggested in the original paper by (Ioffe and Szegedy) . We applied Dropout as another regularization technique, which randomly sets the values of some input neurons to zero (Srivastava, Hinton and Krizhevsky). Finally, we added a L2 regularization (Krogh and Hertz) term in the cross-entropy cost (Mannor, Peleg and Rubinstein) function $J$ as shown in equation(1), which ensures small values of all weight parameters ($\Theta_w$) to prevent the domination of a single weight parameter on the decision of the classifier. The equation contains two terms. The first term is the usual cross-entropy cost which is used to penalize the optimization algorithm, if it predicts $g_i$ instead of the true label $y_i$ for the $i_{th}$ training example $x_i$. The second term represents the sum of the square of all the weights in the above defined architecture, also referred to as L2 regularization. This term is scaled by a factor $\frac{\lambda}{2n}$, where $\lambda$ is a (positive) hyperparameter and $n$ is the mini-batch size.

$$J(\Theta) = \frac{-1}{n}\sum_{i=1}^{n}[y_i ln(g_i) + (1-y_i)ln(1-g_i)] + \frac{\lambda}{2n}\sum_{\Theta_w}\Theta_w^2 \qquad (1)$$

We have used the mini-batch gradient descent with momentum (Qian) as an optimization technique to update the values of weights and biases. The mini-batch gradient decent with momentum considers $n$ training examples to compute a moving average of the gradients (see equation (2)) and then update the weights and biases in a single iteration ($\Theta_j$ represents $j^{th}$ learnable parameter). The required derivatives were calculated by employing the backpropagation algorithm (Rumelhart, Hinton and Williams). The term $\gamma$ in equation (3) is referred as momentum, which is also a hyperparameter.

$$V^t = \gamma V^{(t-1)} + (1-\gamma)\left(\frac{1}{n}\sum_{m=1}^{n}\frac{\partial j(\Theta)}{\partial \Theta_j^m}\right) \qquad (2)$$

$$\Theta_j := \Theta_j - \alpha V^t \qquad (3)$$

### 2.8.3. Tuning of hyperparameters

The learning rate $\alpha$ in the mini-batch gradient descent with momentum (SGDM, equation (3)) started at $\alpha = 0.1$ and got tuned in piecewise manner, decreasing by a factor of 10 every 5 training epochs. The momentum $\gamma$ in SGDM (equation (2)) was selected to be 0.9, so that the algorithm considered the last 10 iterations to calculate the moving average $V^t$ of the gradients (equation (3)). Besides that, the mini-batch size $n$ in equation (2) depends on the available GPU-memory. We used $n = 256$, which was the optimized value for our hardware.

We have tuned the parameter $\lambda$ of L2 regularization (see equation 1) with grid search by defining the grid search space from 0 to 5, with a step size of 0.2, 1.8 was found to be the optimized value. We have also used the early stopping criteria to avoid overfitting by monitoring the validation error on validation data $D_{va}$, at each epoch. If the validation error of six consecutive epochs increased or remained same, the training terminated. Lastly, dropout regularization is used with the probability of dropping input neuron is 0.5.

We have compared the classification accuracy of the USC with a fully connected neural network.

### 2.9. The Architecture of Fully Connected Neural Network

The feature matrix $x \in R^{b \times w}$ was first reshaped into a vector $x \in R^{(b \times w) \times 1}$ (see Figure 5). Here, the batch size $b$ was considered 20 and $w$=48, resulting in a feature vector $x \in R^{960 \times 1}$. It was then propagated forward from input to output layer (see Figure 5). The output of the preceding layer becomes the input of the following layer, and it continues through all three hidden layers and eventually the output layer. The number of neurons in the hidden layers are 500, 250 and 125, and the output layer contains two neurons.

We have used Rectified Linear Units (ReLUs) $f(x) = max(x, 0)$ as an activation function (Nair and Hinton) and Softmax classifier at the last layer (Kaibo Duan).

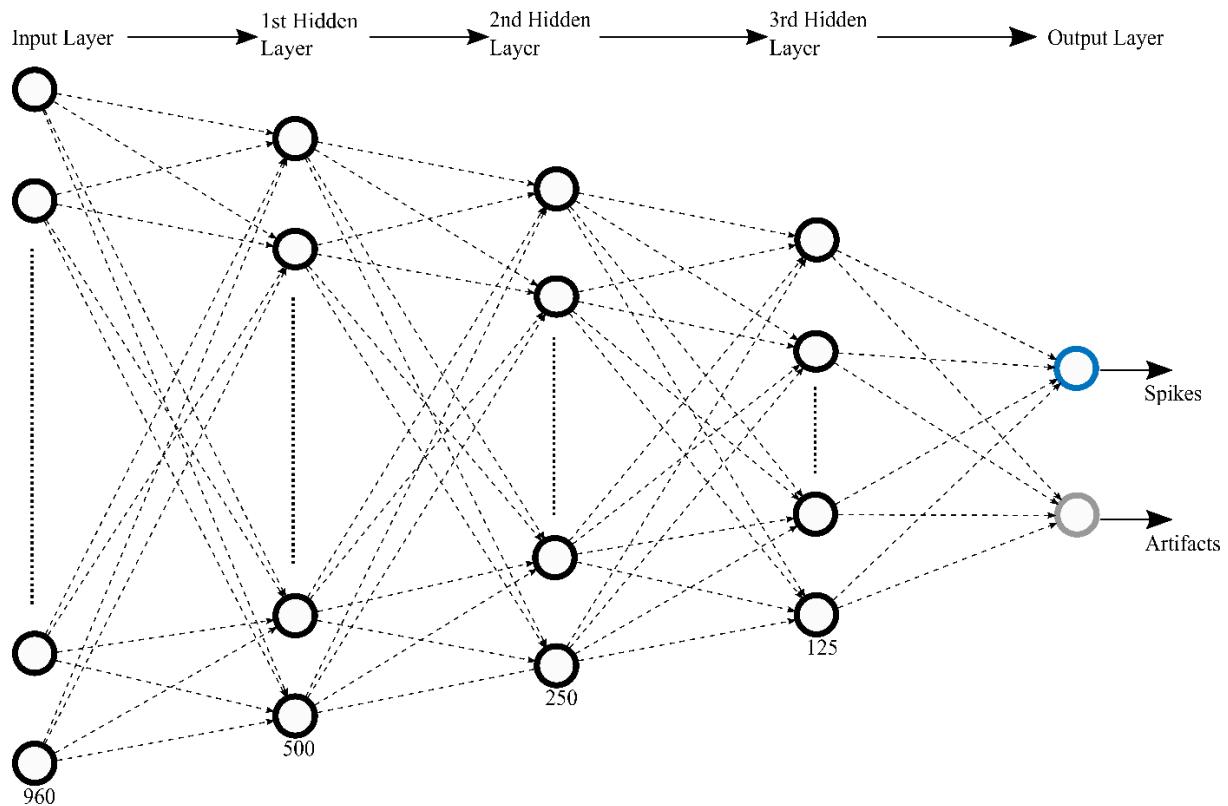

*Figure 5: Architecture of fully connected neural network. The input is processed through three hidden layer and output layer, number of the neurons in each layer are 500,250,125 and 2. Neurons of each layers are fully connected to the neurons of the preceding layer.*

We have used the same regularization and optimization techniques as explained in section (Regularization techniques and optimization algorithm).

We have tuned the hyperparameters of the above defined architecture, similarly the way described in section (Tuning of hyperparameters). However, the optimized value of L2 regularization was found to be 2.4 using grid search method.

## 3. Results
### 3.1. Training and Evaluation

For training the algorithm, we have compiled a dataset from the first six recording sessions of subject U1. The distribution of training data and the exact number of feature vectors casted from each patient is explained in section Data distribution for training algorithm and evaluation of generalization. We have trained the universal spike classifier and the fully connected neural network on the $D_{tr}$ (training data). The training and the validation loss (regularized cross-entropy) of both models were monitored during training on $D_{tr}$ and $D_{va}$ (see Figure 6), respectively. The process of training was terminated once the validation error stopped decreasing or remained unchanged for six consecutive epochs. During training, the CNNs achieved minimum validation error (0.027) at 14[th] epoch as shown in Figure 6(a). After that, there is a rise in validation error and then it remains approximately constant. Finally, the process of training terminated at the 20[th] epoch. Similarly, the FNN achieved its minimum validation error (0.071) at the 17[th] epoch and the process of training was terminated at 23[rd] epochs as shown in Figure 6(b). The values of parameters (weights & biases) were saved at lowest validation error and were used later used to map test inputs to decision space.

The architecture of CNNs is explained in the section **Input Representation and the Architecture of USC** and the architecture of FNN is explained in the section **The Architecture of Fully Connected Neural Network**. The process of optimizing parameters of defined architecture is explained in the section **Regularization techniques and optimization algorithm** and the process of tuning hyperparameters of optimization algorithm and the tuning of hyperparameters of regularization algorithms are explained in the section **Tuning of hyperparameters**.

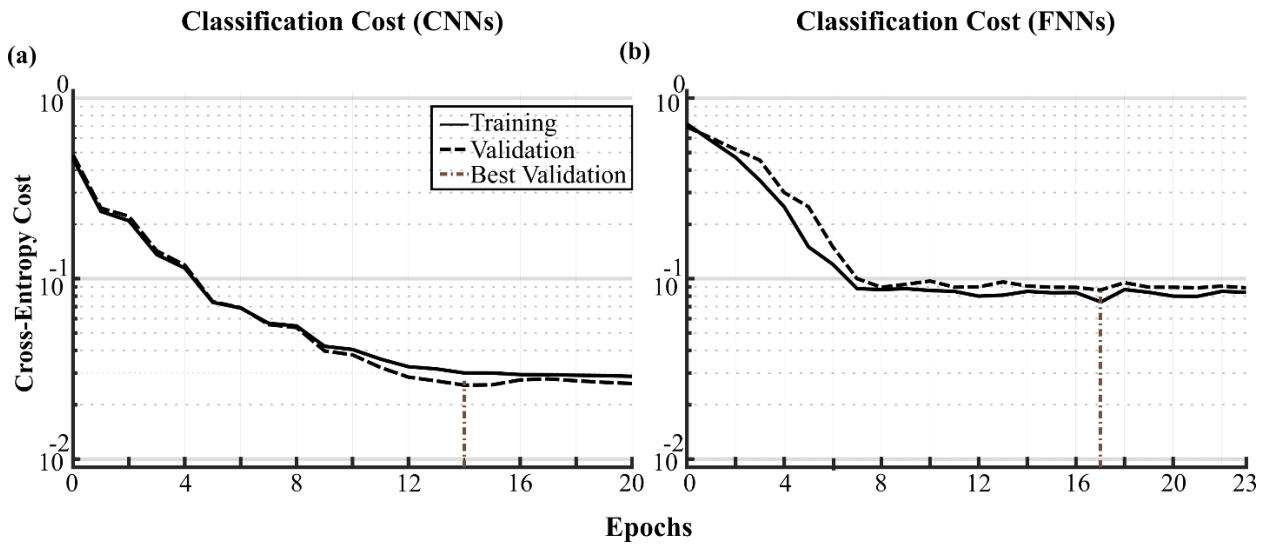

*Figure 6: (a)Training and validation cost of CNNs. (b) Training and validation cost of FNNs.*

## 3.2. Evaluation of Generalization

The generalization quality of the trained models was evaluated using data from eight patients that was remained unseen during training. These patients were implanted with either Utah arrays or microwires, targeting different brain structures and performing different types of behavioral tasks under various experimental and recording conditions. The performance of the trained classifiers on the data collected from patients implanted with Utah arrays and microwires are shown separately in Figure 7, and Figure 8, respectively. The data distribution of the patients implanted with Utah arrays is highly unbalanced, 34.4% data represent spikes and remaining 65.6% data represent artifacts as shown in confusion matrix (see Figure 7). Therefore, the evaluation accuracy of each individual class is much more important, than cumulative accuracy. Main goal of the classifier is to detect and track the channel containing neural data. Hence, accuracy of CNNs and FNNs on the feature vectors representing spikes plays a pivotal role, which is 94.6% and 92.6%, respectively. There is a significant difference of 2% in the evaluation performance of both classifiers, CNNs detect the channels with spike data more efficiently. Moreover, the evaluation performance of CNNs and FNNs on the feature vector representing artifacts is 97.9% and 98.6%, respectively. Similarly, the overall accuracy of CNNs and FNNs is 96.7 % and 96.5% as shown in Figure 7.

**Classification Accuracy (Utah Array)**

|  | CNNs | | | | FNNs | | |
|---|---|---|---|---|---|---|---|
| **Spikes** | 388317 / 32.5% | 16573 / 1.4% | 95.9% | **Spikes** | 380463 / 31.9% | 11266 / 0.9% | 97.1% |
| **Artifacts** | 22282 / 1.9% | 766528 / 64.2% | 97.2% | **Artifacts** | 30136 / 2.5% | 771835 / 64.7% | 96.2% |
|  | **94.6%** | 97.9% | **96.7%** |  | **92.6%** | 98.6% | **96.5%** |
|  | Spikes | Artifacts |  |  | Spikes | Artifacts |  |

(Predicted Labels / True Labels)

*Figure 7: Classification accuracy of trained Models (CNNs and FNNs) on the data collected from patients implanted with Utah array, individually.*

The data distribution of patients implanted with microwires is even more unbalanced, only 1.65% data represent feature vectors of class 'spikes' and remaining 98.35% data represent feature vectors of class 'artifacts' (see Figure 8). Accuracy of CNNs and FNNs on the feature vectors representing spikes is 97.2% and 91.9%. Here, CNNs outperforms the FNNs with the significant difference in evaluation performance of 5.3%. The evaluation performance of CNNs and FNNs on the data represents the feature vectors of class 'artifacts' is 98.9% and 98.6%. The overall accuracy of CNNs and FNNs is 98.9 % and 98.6% as shown Figure 8.

The results show in (Figure 7) and (Figure 8) provide enough that CNNs outperforms FNNs generally, as well as on those where neural data is present.

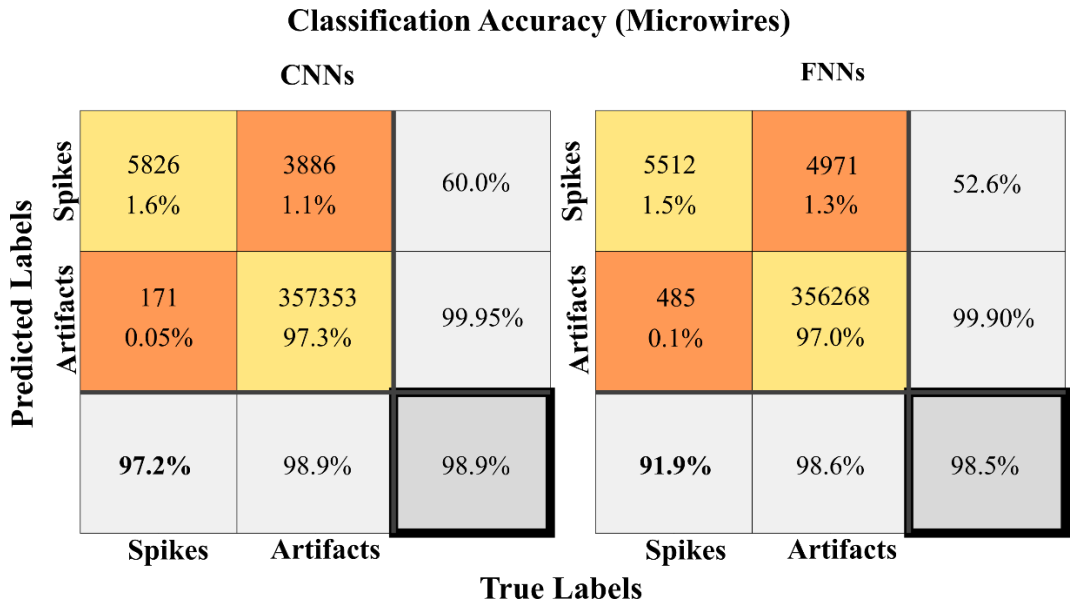

*Figure 8: Classification accuracy of trained models (CNNs and FNNs) on the data collected from all patients implanted with Microwires, individually.*

## 3.3. Impact of Batch Size

The box plots in Figure 9 show the classification accuracy on the test data, when the USC is trained and evaluated for ten times at the corresponding batch size, and the time on x-axis is the average time of all channels to construct a feature vector of the corresponding batch size. The classification accuracy increases with increasing the batch size (see Figure 9). However, by increasing the batch size construction of feature vector takes more time, and by decreasing batch size the classification accuracy drops as shown in Figure 9. Therefore, this tradeoff between the classification accuracy and choosing the right batch size needed to be optimized. The classification accuracy with batch size 20 was 97.5% and reached 99.5% with batch size 65 before it got saturated. The time to construct a single vector is much more critical, when we are using USC for meaningful channel selection for feature extraction in online decoding. The time to construct a feature vector with batch size = 10 is 280ms, which provide acceptable classification of 97%. Therefore, it is possible construct a feature vector and track neural data from each channel, during online decoding. On the other side, if we are using USC in offline spike sorting for meaningful channel selection, we can construct a feature vector with batch size = 65 because after that classification accuracy saturates at 99.5 % but time to construct keep increasing.

Here, we have chosen batch size=20 for all the analysis of this research work. However, the selection of batch size depends on the application.

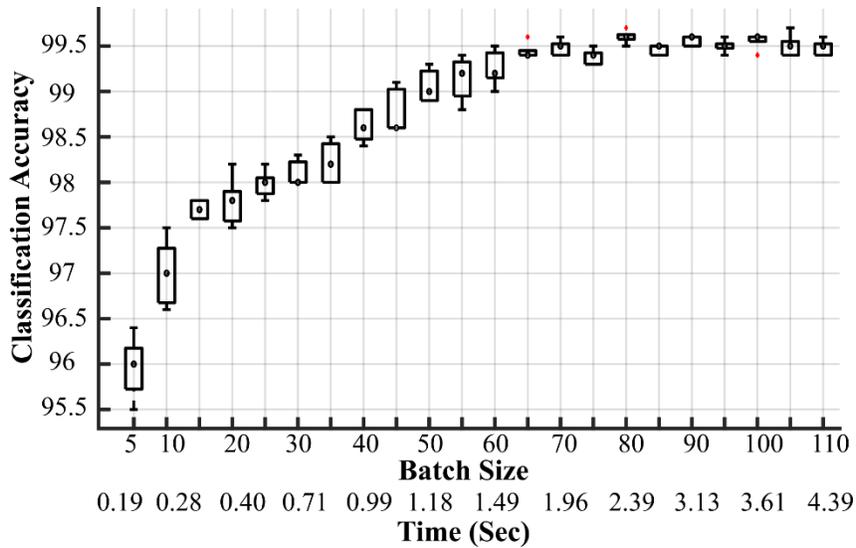

*Figure 9: Impact of Batch size on classification accuracy and time to construct feature vector.*

## 3.4. Tracking of Neural Data

The neural data can be lost, or neural data appeared on some channels with the passage of time or even during the same recording session. Therefore, it is very important to track the presence or absence of neural data from the channels. Another important aspect of this research work is to provide a way to track the presence or absence of neural unit on the channel at runtime. We supported this claim by employing USC on the data of three different kinds of channels (see Figure 10), where the presence of a unit is stable, less stable or unstable. A fluctuation between presence and absence of neural units takes place occasionally in partially stable channels and much more frequently in unstable channels (shown in Figure 10). On stable channels a unit is either present or absent during one complete recording session, as shown in Figure 10. This result provides evidence that the USC tracks the presence or absence of units comprehensively on all types of channels.

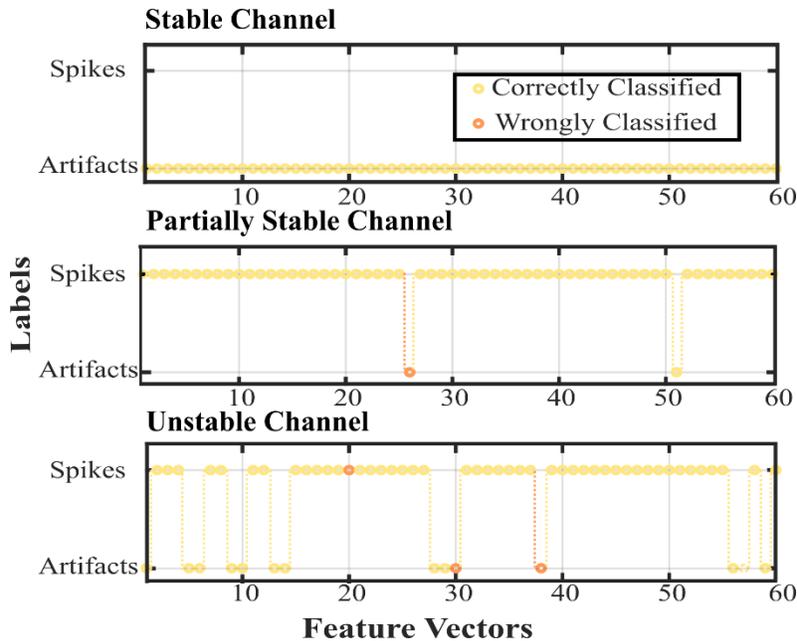

*Figure 10: Performance of USC on tracking of neural data on different types of channels (stable, partially stable, unstable)*

## 3.5. Visualization of correctly and wrongly classified examples

We have selected a small subset of correctly and wrongly classified input for the sake of visualization and evaluation of the performance of the USC. The examples were selected randomly from all 136 recording sessions and are shown in Figure 11 and Figure 12.

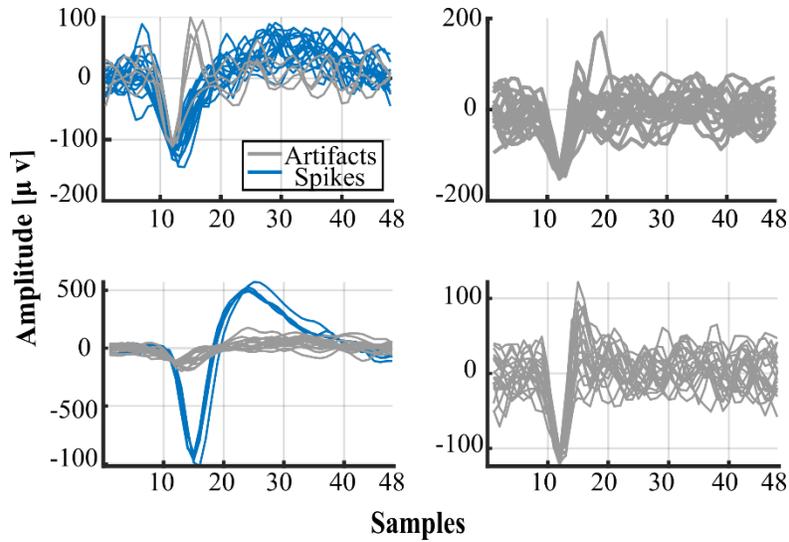

*Figure 11: Randomly selected correctly classified examples. The events representing spikes are shown in blue and events presenting artifacts are shown in grey color.*

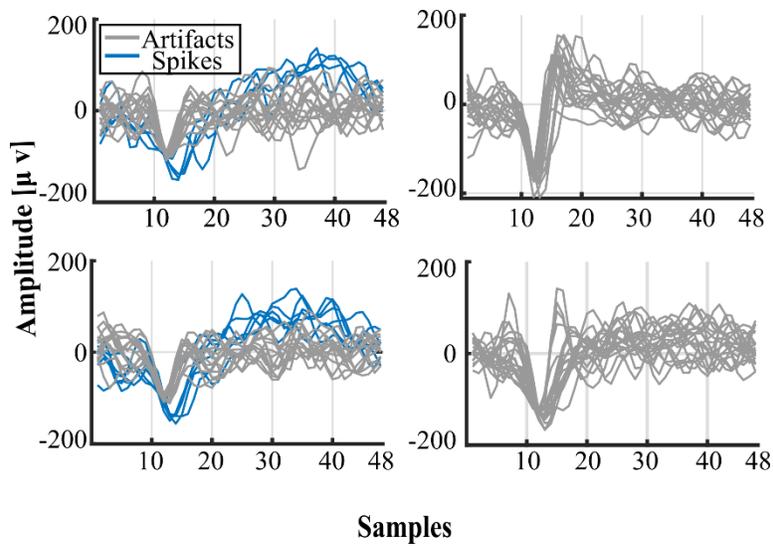

*Figure 12: Randomly selected wrongly classified examples. The events representing spikes are shown in green and events presenting artifacts are shown in red color.*

## 4. Discussion

In this study, we developed a novel spike classification algorithm, the universal spike classifier, which discriminates channels recording single cell data from channels recording only noise. This discrimination works with data collected using different subjects, brain areas, recording sessions and different types of recording microelectrodes. The evaluation performance of CNNs and FNNs on the data collected from eight subjects (see Figure 7 & Figure 8) provides evidence that the trained model adapts the generalization properly. We also showed that the CNNs outperform the FNNs on channels where units are present (see Figure 7 & Figure 8) on both types of implanted electrodes, and performed equally well on the channels, where neural data is absent.

We have also shown from some of the correctly and the wrongly classified examples (see Figure 11 & Figure 12). The correctly classified examples show that USC performed very well on accepting and rejecting channels. However, some of the wrongly classified examples are very difficult, even human expert can disagree on the decision of assigning labels to some individual events, e.g., event representing spikes in the example at position (1,1), and some of the events representing artifacts in the example at position (2,2) in the Figure 12.

Our USC has several useful applications. During online decoding in invasive brain computer interface (BCI), feature vectors were previously constructed from threshold crossing events (Klaes, Kellis and Afalo; W Fraser, M Chase and Whitford; Koyama, Chase and Whitford; Aflalo, S and Klaes). With the USC threshold crossing events not representing single unit activity can be easily discarded. As a result, more meaningful feature vectors can be constructed. Moreover, previously, it has also been shown in BCI studies that units disappear, or new units appear with the passage of time (Lebedev; A. Lebedev; Moritz and Fetz; Sanchez, Carmena and Lebedev).By employing USC, one can automatically track the presence or absence of single unit activity at runtime (see (Tracking of Neural Data)).

Most existing spike sorting methods operate either manual or semi-automatic (Einevoll GT; Lewicki; Olivier Marre; Quiroga). In manual spike sorting, user curation is required for channel selection and for clustering. The latter is done by attending to the features in a two-dimensional projection and defining the cluster centers and boundaries manually. Here, USC can be employed to reduce the time and effort needed at the first stage by automatically discarding the channels that contain only artifacts. In semi-automatic spike sorting, clustering is done automatically, but the user must curate in order to decide which cluster to reject and which cluster to accept (Daniel N. Hill; Kadir SN; Rossant C). Contrarily, the USC can either accept or reject the cluster automatically, by constructing the feature vector from the waveforms of corresponding cluster (as explained in (Batch size)).

It is possible that a channel records MUA. The source separation (SUA) is another important issue that needs to be addressed. However, the current version of USC does not address this issue. Although many methods exist to solve this issue, very few of them offer fully automated spike sorting (Chung JE; Grossberge, P. Battaglia and Vinck; Hossein Nadian, Karimimehr and Doostmohammadi) and none of them offer a universal fully automated spike sorter. Therefore, in future we aim to extend our USC to a universal spike detector (USD) that will not only determine the presence or absence of neural data on the channel, but also detect and track every present neural source, universally.

The current version of USC detects and track channels containing neural data recorded from humans, only. In the future we also want to extend the scope of USC to data from different type of spikes (excitatory, inhibitory) (Andrea Becchetti) and other species (rat, cat) and different types of neural cells.

# Bibliography

A. Lebedev, Mikhail. „How to read neuron-dropping curves?" *Frontier in Systems Neuroscience* (2014): 8:102.

Abeles, M und M.H Goldstein. „Multispike train analysis." *Proceedings of the IEEE* (1977): 762-773.

Aflalo, T, et al. „Neurophysiology. Decoding motor imagery from the posterior parietal cortex of a tetraplegic human." *Science* (2015): 348(6237):906-10.

Andrea Becchetti, Francesca Gullo, Giuseppe Bruno, Elena Dossi, Marzia Lecchi, Enzo Wanke. „Exact distinction of excitatory and inhibitory neurons in neural networks: a study with GFP-GAD67 neurons optically and electrophysiologically recognized on multielectrode arrays." *Frontiers in Neural Circuits* (2012): 6, 63 .

B Schwartz, A. „Cortical neural prosthetics." *Annual Review of Neuroscience* (2004): 487-507.

Ballini, Marco, et al. „A 1024-Channel CMOS Microelectrode Array With 26,400 Electrodes for Recording and Stimulation of Electrogenic Cells In Vitro." *IEEE Journal of Solid-State Circuits* (2014): 2705-2719.

Berényi, A, et al. „Large-scale, high-density (up to 512 channels) recording of local circuits in behaving animals." *J Neurophysiology* (2013): 1132-49.

Chung JE, Magland JF, Barnett AH, Tolosa VM, Tooker AC, Lee KY, Shah KG, Felix SH, Frank LM, Greengard LF. „A Fully Automated Approach to Spike Sorting." *Neuron* (2017).

Chung, Jason E., et al. „A Fully Automated Approach to Spike Sorting." *Neuron* (2017).

Daniel N. Hill, Samar B. Mehta, and David Kleinfeld. „Quality metrics to accompany spike sorting of extracellular signals." *The Journal of neuroscience : the official journal of the Society for Neuroscience* (2011): 8699-8705.

Deng, Jia, et al. „ImageNet: A large-scale hierarchical image database." *Computer Vision and Pattern Recognition, 2009*. Miami, FL, USA: IEEE, 2009.

Dragas, Jelena, et al. „A Multi-Functional Microelectrode Array Featuring 59760 Electrodes, 2048 Electrophysiology Channels, Stimulation, Impedance Measurement and Neurotransmitter Detection Channels." *IEEE journal of solid-state circuits* (2017): 1576-1590.

Einevoll GT, Franke F, Hagen E, Pouzat C, Harris KD. „Towards reliable spike-train recordings from thousands of neurons with multielectrodes." *Current Opinion in Neurobiology* (2012): 11-17.

Frey, U, et al. „Microelectronic system for high-resolution mapping of extracellular electric fields applied to brain slices." *Biosensors & bioelectronics* (2008): 2191-8.

Fried I, Wilson CL, Maidment NT, Engel J Jr, Behnke E, Fields TA, MacDonald KA, Morrow JW, Ackerson L. „Cerebral microdialysis combined with single-neuron and electroencephalographic recording in neurosurgical patients. Technical note." *J Neurosurg.* (1999 ): 697-705.

Fried, et al. „Cerebral microdialysis combined with single-neuron and electroencephalographic recording in neurosurgical patients. Technical note." *J Neurosurg.* (1999): 697-705.

Gibson, Sarah, Jack W. Judy und Dejan Marković. „Spike Sorting: The First Step in Decoding the Brain: The first step in decoding the brain." *IEEE Signal Processing Magazine* (2012): 124-143.

Grossberge, Lukas, Francesco P. Battaglia und Martin Vinck. „Unsupervised clustering of temporal patterns in high-dimensional neuronal ensembles using a novel dissimilarity measure." *PLOS Computational Biology* (2018): l 14(7): e1006283.

Guo, Tianmei, et al. *Big Data Analysis (ICBDA), 2017 IEEE 2nd International Conference on*. Beijing, China: IEEE, 2017.

Guo, Tianmei, Jiwen Dong und Henjian Li. „Simple convolutional neural network on image classification." *IEEE 2nd International Conference on Big Data Analysis (ICBDA)*. Beijing: IEEE, 2017.

Herculano-Houzel, Suzana. „The Human Brain in Numbers: A Linearly Scaled-up Primate Brain." *Frontiers in Human Neuroscience* (2009): 3:31.

Hill, Daniel N., Samar B. Mehta und David Kleinfeld. „Quality Metrics to Accompany Spike Sorting of Extracellular Signals." *The Journal of Neuroscience* (2011): 8699-8705.

Hodgkin, A.L und A.F Huxley. „A Quantitative Description of Membrane Current and its Application to Conduction and Excitation in Nerve." *Journal of Physiology* (1952): 500-544.

Hossein Nadian, Mohammad, et al. „A fully automated spike sorting algorithm using t-distributed neighbor embedding and density based clustering." *bioRxiv* (2018): bioRxiv 418913.

Ioffe, Sergey und Christian Szegedy. „Batch Normalization: Accelerating Deep Network Training by Reducing." *International Conference on Machine Learning*. Lille, 2015.

Kadir SN, Goodman DF, Harris KD. „High-dimensional cluster analysis with the masked EM algorithm." *Neural Computation* (2014): 2379-94.

Kaibo Duan, S. Sathiya Keerthi, Wei Chu, Shirish Krishnaj Shevade, Aun Neow Poo. „Multi-category classification by soft-max combination of binary classifiers." *MCS'03 Proceedings of the 4th international conference on Multiple classifier systems*. Berlin: Springer, 2003. 125-134.

Kita, Justin M. und R. Mark Wightman. „Microelectrodes for Studying Neurobiology." *Current opinion in chemical biology* (2008): 491–496.

Klaes, Chrsitian, et al. „Hand Shape Representations in the Human Posterior Parietal Cortex." *The Journal of Neuroscience* (2015): 15466-15476.

Kotsiantis, S. B. „Supervised Machine Learning: A Review of Classification ." *Informatica 31* (2007): 249-268.

Koyama, S, et al. „Comparison of brain-computer interface decoding algorithms in open-loop and closed-loop control." *Journal Of Computaional Neuroscience* (2009): 73-87.

Krizhevsky, Alex, Ilya Sutskever und Geoffrey E Hinton. „ImageNet Classification with Deep Convolutional." *Neural Information Processing Systems (NIPS)* (2012): 1097-1105.

Krizhevsky, Alex, Ilya Sutskever und Geoffrey E. Hinton. „ImageNet Classification with Deep Convolutional Neural Networks." *Neural Information Processing Systems (NIPS)*. 2012.

Krizhevsky, Alex, Ilya Sutskever und Geoffrey Hinton. 2012. 1097-1105.

Krogh, Anders und John A. Hertz. „A Simple Weight Decay Can Improve Generalization." *Neural Information Processing Systems (NIPS)*. Neural Information Processing Systems (NIPS), 1991.

Lambacher, A, et al. „Identifying firing mammalian neurons in networks with high-resolution multi-transistor array (MTA)." *Applied Physics A* (2011): 1-11.

Lebedev, Mikhail A. „How to read neuron-dropping curves?" *Frontiers System Neuroscience* (2014): 8:102.

LeCun, Yann, et al. „Gradient-based learning applied to document recognition." *IEEE*. IEEE, 1998. 2278-2324.

Lewicki, Michael S. „A review of methods for spike sorting: the detection and classification of neural action potentials." *Network 9* (1998): R53-R78.

Mannor, Shie, Dori Peleg und Reuven Rubinstein. „The cross entropy method for classification." *ICML '05 Proceedings of the 22nd international conference on Machine learning*. Bonn, 2005. 561-568.

Matthews BA, Clements MA. „Spike Sorting by Joint Probabilistic Modeling of Neural Spike Trains and Waveforms." *Computational Intelligence and Neuroscience* (2014).

Moritz, CT und EE Fetz. „Volitional control of single cortical neurons in a brain-machine interface." *Journal of Neural Engineering* (2011): 2:025017.

N.Hill, Daniel, Samar B.Mehta und David Kleinfeld. „Quality metrics to accompany spike sorting of extracellular signals." *Journal Of Neuro* (2011): 8699-8705.

Nair, Vinod und Geoffrey E. Hinton. „Rectified Linear Units Improve Restricted Boltzmann Machines." *International Conference on Machine Learning, *. Haifa, 2010.

Olivier Marre, corresponding author, Dario Amodei, Nikhil Deshmukh, Kolia Sadeghi, Frederick Soo, Timothy E. Holy, and Michael J. Berry. „Mapping a Complete Neural Population in the Retina." *The Journal of Neuroscience* (2012): 14859-14873.

Pachitariu, Marius, et al. „Fast and accurate spike sorting of high-channel count probes with KiloSort." *Neural Information Processing Systems (NIPS)*. 2016.

Qian, Ning. „On the momentum term in gradient descent learning algorithms." *Neural Networks* (1999): 145-151.

Quiroga, Rodrigo Quian. „Concept cells: the building blocks of declarative memory functions." *Nature Reviews Neuroscience* (2012): 587-597.

Rey, Gonzalo Hernan, Carlos Pedreira und RodrigoQuian Quiroga. „Past, present and future of spike sorting techniques." *Brain Research Bulletin* (2015): 106-117.

Rossant C, Kadir SN, Goodman DFM, Schulman J, Hunter MLD, Saleem AB, Grosmark A, Belluscio M, Denfield GH, Ecker AS, Tolias AS, Solomon S, Buzsaki G, Carandini M, Harris KD. „Spike sorting for large, dense electrode arrays." *Nature Neuroscience* (2016): 634-641.

Rumelhart, David E., Geoffrey E. Hinton und Ronald J. Williams. „Learning representations by back-propagating errors." *Nature* (1986): 533-536.

S Lewicki, Michael. „A review of methods for spike sorting: the detection and classification of neural action potentials." *Network: Computation in Neural Systems* (1998): R53-R78.

Sanchez, JC, et al. „Ascertaining the importance of neurons to develop better brain-machine interfaces." *IEEE Transactions on Bio-medical Engineering* (2004): 943-53.

Sonia Todorova, Patrick Sadtler, Aaron Batista, Steven Chase, Valérie Ventura. „To sort or not to sort: the impact of spike-sorting on neural decoding performance." *Journal of Neural Engineering* (2014).

Spira, ME und A Hai. „Multi-electrode array technologies for neuroscience and cardiology." *Nature Nanotechnology* (2013): 83-94.

Srivastava, Nitish, et al. „Dropout: A Simple Way to Prevent Neural Networks from Overfitting." *Journal of Machine Learning Research 15* (2014): 1929-1958.

Stallkamp, Johannes, et al. „The German Traffic Sign Recognition Benchmark: A multi-class classification competition." *Neural Networks (IJCNN), The 2011 International Joint Conference on Neural Networks*. San Jose, CA, USA: IEEE, 2011.

Tsai, David, et al. „A very large-scale microelectrode array for cellular-resolution electrophysiology." *Nature Communications* (2017).

W Fraser, George, et al. „Control of a brain–computer interface without spike sorting." *Journal of Neural Engineering* (2009): Volume 6, Number 5.